\documentclass{aipproc}

\newcommand{\tbk}{\ensuremath{\tau_{\mbox{bk\ }} }}
\newcommand{\tbki}{\ensuremath{\tau_{\mbox{bk1\ }} }}
\newcommand{\tbkii}{\ensuremath{\tau_{\mbox{bk2\ }} }}
\newcommand{\tbu}{\ensuremath{\tau_{\mbox{bu\ }} }}
\newcommand{\tel}{\ensuremath{\tau_{\mbox{el\ }} }}
\newcommand{\tend}{\ensuremath{t_{\mbox{end\ }} }}
\newcommand{\tskip}{\ensuremath{\tau_{\mbox{skip\ }} }}
\newcommand{\nbki}{\ensuremath{n_{\mbox{bk1\ }} }}
\newcommand{\nbkii}{\ensuremath{n_{\mbox{bk2\ }} }}
\newcommand{\nbu}{\ensuremath{n_{\mbox{bu\ }} }}
\newcommand{\mubu}{\ensuremath{\mu_{\mbox{bu\ }} }}
\newcommand{\ssbk}{\ensuremath{\Sigma^2\ }}
\newcommand{\ssnbu}{\ensuremath{\sigma_{\mbox{bu\ }}^2 }}

\layoutstyle{8x11double}

\title{A ``Spiffy'' Trigger for Gamma-Ray Bursts}

\author{Carlo Graziani}
{
address={
Department of Astronomy \& Astrophysics, University of Chicago,
5640 South Ellis Avenue, Chicago, IL 60637
},
email={carlo@oddjob.uchicago.edu},
}

\begin{abstract}

The traditional design of trigger algorithms for GRB experiments requires
the specification of the background and burst samples in terms of
acquisition times that are of fixed duration and of fixed elapsed time from
each other.  One such set of acquisition times is required for each
characteristic timescale of GRB variation that one desires to detect.  One
then slides each set through the trigger data searching for samples that
maximize the signal-to-noise of the background-subtracted burst sample.

Here we describe a new triggering approach in which the times at which the
background and burst samples are acquired are allowed to vary
dynamically.  Two background samples bracket a burst sample. The
background and burst durations and elapsed time between them are allowed
to be free parameters, which are maximized using the downhill simplex
method.  This produces great flexibility in the timescales that are
available for detecting GRBs.

\end{abstract}

\begin{document}

\maketitle

\section{Introduction}

The search for untriggered GRBs is has been an active field of research at
least since the public release of BATSE data.  Most recently, Kommers et
al. \citep{kommers01} and Stern et al. \citep{stern01} have described
searches of BATSE data directed at revealing GRBs that occurred without
being detected by the BATSE on-board triggers, either for operational
reasons or because their spectral or temporal morphologies were poor fits
to the on-board trigger criteria.

Naturally, the same considerations apply to GRB detection by HETE.  The
HETE mission deploys an unprecedentedly varied set of trigger criteria ---
the FREGATE DSP trigger \citep{atteia02} uses four timescales and operates
in two energy bands, while the WXM XG trigger
\citep{fenimore01,tavenner02} is typically configured to apply thirty or
so criteria, some on WXM data and others on FREGATE data.  Nevertheless
the variety of GRB morphologies, and operational considerations, can result
in GRBs that are not detected in flight.  It is important to develop a
strategy to mine HETE survey data for such untriggered GRBs.

Typical ground searches for untriggered bursts use detection methods that
largely mirror on-board trigger algorithms \citep{kommers01,stern01}. 
Background and burst samples are specified in terms of acquisition times
that are of fixed duration and of fixed elapsed time from each other. 
Each GRB timescale --- risetime or duration --- is probed by a different
fixed choice of these parameters.  Each such fixed set of time windows is
then swept through the time series being probed for transient events,
searching for samples that maximize the signal-to-noise of the
background-subtracted burst sample.

This scheme has the disadvantage of being rather inflexible about the the
timescales that are probed.  This inflexibility is especially troublesome
when seeking weak signals, for which inaptly chosen burst or background
samples may lead to a signal dilution that prevents detection.

In this work we describe an alternative approach that has been quite
successful in identifying extremely weak events.  In this approach, the
background and burst samples are treated as free parameters, which are
varied using the downhill simplex method of Nelder \& Mead
\citep{nelder65} to maximize the signal-to-noise ratio of the
background-subtracted burst sample.

\section{Implementation}

\begin{figure}[ht]
\includegraphics[scale=0.3,angle=-90]{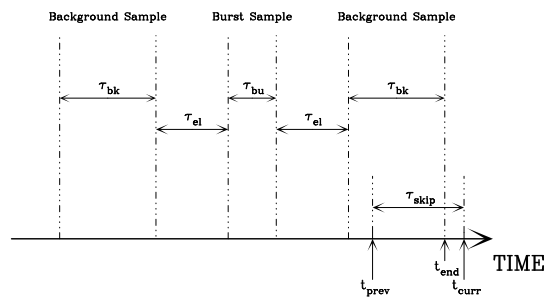}
\caption{
Illustration of the definition of the time parameters that are varied to
search for transient events.
}
\label{timefig}
\end{figure}

The operation of the code, {\tt spiffy-trigger}, is illustrated in Figure
\ref{timefig}.

The trigger operates as a simplified ``bracket trigger''
\citep{fenimore01,tavenner02}, in that the background is estimated using
samples before and after the burst sample.  The background is assumed
constant, so no interpolation (linear or otherwise) is performed to obtain
the background rate during the burst sample.

This restriction is not an essential feature of the method, but rather
merely a simplification. The two background intervals are restricted to
remain equidistant from the burst sample, so as to prevent the
maximization procedure from exploiting a monotonic increase or decrease in
the background rate to estimate an erroneously low background, by driving
one of the background samples to a region of lower background without
driving the other to a region of higher background.

The code operates on a time-series of integer counts.  It advances a
trigger window of fixed duration through the time series by steps of size
\tskip.  It sets up a burst sample interval, of duration \tbu, bracketed
at an elapsed time {\tel} by two background sample intervals of duration
\tbk, the second of which ends at time \tend.  It calculates the SNR for
the burst sample, assuming a background rate calculated by a weighted
average of the count rates in the two background intervals.

The SNR is computed as follows:  Assume for the sake of generality that
the two background accumulation times may differ, so that we accumulate
\nbki counts in the first background during an accumulation time \tbki,
and \nbkii counts in the later background accumulation time \tbkii. 
Denoting the estimated background counts during the burst sample by \mubu,
and assuming the Gaussian approximation to the Poisson distribution,
it is a straightforward exercise in Gaussian estimation to show that
\begin{eqnarray}
\mubu&=&\frac{\tbki+\tbkii}{\tbu}\,\ssbk,\\
\ssbk&=&\tbu^2
  \left(\frac{\tbki^2}{\nbki}+\frac{\tbkii^2}{\nbkii}\right)^{-1},
\end{eqnarray}
where \ssbk is the variance in the estimate \mubu.

Denote by \nbu the counts that we accumulate during the burst sample. 
Then the net signal in the burst sample is $s=\nbu-\mubu$.  The variance
in $s$ is the sum of \ssbk and the variance in \nbu.  Triggering is
essentially hypothesis testing, with the null hypothesis consisting of the
assumption that the count rate in the burst sample is the same as what is
estimated using the background samples.  Thus the appropriate choice for
the variance of \nbu is ``model variance'', that is $\ssnbu=\mubu$.  Thus
the SNR of the burst sample is
\begin{equation}
\mbox{SNR}=\frac{\nbu-\mubu}{\left(\mubu+\ssbk\right)^{1/2}}.
\label{snreq}
\end{equation}
This is the quantity that {\tt spiffy-trigger} endeavors to maximize.

The code uses the simplex method to vary the four parameters \tend, \tbk,
\tel, and \tbu, which are viewed by the simplex minimization routine as
continuous parameters.  A very lax convergence criterion is imposed ---
the absolute variation of the SNR must be less than 0.1 across the simplex
--- because in triggering there is no point in determining the SNR to
great accuracy, and because we don't want to spend many CPU cycles chasing
noise.

The parameter \tend  is constrained to be later than the end of the trigger
window in the previous invocation.  Consequently, the arrangement of burst
and background samples ``accordions out'' backwards in time from the
current time, without repeating choices of intervals made during previous
iterations. 

When there is no transient event in the data, the simplex will typically
not wander very far from its initial configuration.  On the other hand if
there is a transient event, and the initial simplex includes a vertex
corresponding to a configuration in which the burst sample even partially
includes the event, the simplex will rapidly climb the SNR slope,
dynamically adjusting its timescales until the event is well-bracketed.

Since the simplex does not wander far if it doesn't find much at the
outset, it is important to ensure that \tskip is not so large that a short
event may ``fall between the cracks'' --- that is, fail to have any of its
constituent time samples included in a burst sample probed by the initial
simplex. It is therefore a good idea to ensure that at simplex
initialization, $\tbu>\tskip$ for at least one of the simplex vertices. 
This ensures that every data sample passes through the burst sample of at
least one initial simplex parameter vertex.

Constraints on the time parameters are imposed by making the SNR function
return a large negative value when the constraints are violated.  The
previously-discussed constraint on the parameter \tend is enforced in this
way.  The code also uses this parameter-constraint mechanism to prevent
intervals from encroaching upon each other, to ensure that \tbk, \tel, and
\tbu remain positive-valued, and to keep all intervals inside the current
trigger window.

Other useful constraints that it is good practice to enforce are
a minimum value for \tel (so that burst and background samples are
well-separated), a minimum duration for \tbk (so as to minimize the risk
of the background nestling into a low fluctuation), and a maximum duration
\tbu (so as to minimize the risk of triggering on very long duration
trends in the background).

\section{Deployment}

{\tt spiffy-trigger} is currently used in three different contexts within
the HETE project:

\begin{figure}[t]
\includegraphics[scale=0.3,angle=-90]{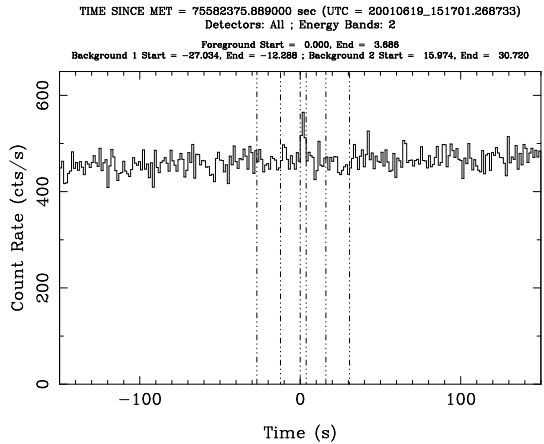}
\caption{
GRB detected by the trigger robot on 19 June 2001 at 15:17:01 UTC.  The
SNR of this event was 6.1, the duration was 3.7 s.  This event was also
detected by BeppoSax.
}
\label{robotgrb}
\end{figure}

\begin{itemize}

\item The Chicago ground location pipeline \citep{graziani02} uses {\tt
spiffy-trigger} to identify the burst sample time with maximal
signal-to-noise in the WXM data.  This sample is used throughout the
subsequent location analysis.

\item A robot script that runs after every downlink uses {\tt
spiffy-trigger} to search for untriggered bursts in FREGATE band C (40-300
keV) 1.3s resolution survey data.  During normal HETE operation, it tends
to see about 1 possible GRB per week, above and beyond detecting all
triggers picked up in flight that are sufficiently hard, and long (or
short but bright) to register at this timescale and in this energy band. 
Figure \ref{robotgrb} shows an example of such an event, which was
confirmed by BeppoSax.

\item The general untriggered burst search described by Butler \& Doty
\citep{butler02} uses {\tt spiffy-trigger} in parallel to Butler \& Doty's
wavelet trigger, and runs on all survey data products.  GRB011212 was in
fact detected on the ground in this pipeline, by both the wavelet algorithm
and by {\tt spiffy-trigger}.

\end{itemize}

\section{Conclusions}

The {\tt spiffy-trigger} algorithm can probe a wide spectrum of burst
timescales.  It is still possible that initialization with a very short
\tbu might miss a very long, slow-rising event, or that a very long
initial \tbu might cause the SNR of a weak, short event to be too diluted
to register before convergence is reached.  However, careful choice of the
range of \tbu spanned by the initial simplex can address this issue to a
large extent.

In any event, the algorithm may be re-run with radically different initial
values of \tbu.  For example, re-running the algorithm three times, with
\tbu set initially to 0.1s, 3s, and 100s --- with suitably chosen initial
simplices --- one may probe a range of timescales that would probably
require hundreds of criteria for a traditional trigger algorithm to
examine.

In principle, there is no reason the {\tt spiffy-trigger} algorithm could
not be deployed in flight in a future mission.  The floating-point
operations that it performs are not particularly expensive, particularly
for modern space computing hardware.  While more complex than a traditional
trigger, it is not vastly more so, and its complexity is offset by its
great flexibility, configurability, and dynamic range of burst timescales
to which it is sensitive.

\end{document}